\documentclass[11pt]{article}
\usepackage{amsmath}
\usepackage{amssymb}
\usepackage{hhline}
\usepackage[scale={.75,.75}]{geometry}
\usepackage{url}
\usepackage{lscape}
\usepackage{caption}
\usepackage[numbers, sort&compress]{natbib}
\usepackage{epsfig}
\usepackage{multirow} 
\usepackage{color}
\usepackage{verbatim}
\usepackage{nicefrac}
\usepackage{upgreek}
\usepackage{bbm}
\usepackage{setspace}
\usepackage{pstricks}
\usepackage{color}

\usepackage{hyperref}% add hypertext capabilities

\newcommand{\im}{{\rm Im}}

\renewcommand{\k}{\textbf{k}}

\newcommand{\p}{\textbf{p}}
\newcommand{\x}{\textbf{x}}
\newcommand{\y}{\textbf{y}}

\newcommand{\tphi}{\tilde{\phi}}
\newcommand{\tchi}{\tilde{\chi}}
\newcommand{\tg}{\tilde{g}}
\newcommand{\rephi}{A}
\newcommand{\imphi}{B}
\newcommand{\rechi}{C}
\newcommand{\imchi}{D}
\definecolor{green}{rgb}{0,0.5,0}

\begin{document}
\date{}

\title{\vspace{-2.5cm} 
%{\normalsize TTK-12-04\hfill\mbox{}\\}
%\vspace{0.5cm}
\begin{flushright}
\vspace{-0.4cm}
{\scriptsize \tt TUM-HEP-914/13}  % 11-18
\end{flushright}
\vspace{-0.3cm}
{\bf Novel collective excitations in a hot scalar field theory}}

\author{Marco Drewes\\
\footnotesize{Physik Department T70, Technische Universit\"at M\"unchen, }\\
\footnotesize{James Franck Stra\ss e 1, D-85748 Garching, Germany}}
%marco.drewes@tum.de}

\maketitle

%\begin{history}
%\received{Day Month Year}
%\revised{Day Month Year}
%%\accepted{Day Month Year}
%%\comby{(xxxxxxxxxx)}
%\end{history}
\vspace{-0.5cm}
\begin{abstract}
%\footnotesize
  \noindent
We study the spectrum of quasiparticles in a scalar quantum field theory at high temperature.
Our results indicate the existence of novel quasiparticles with purely collective origin at low momenta for some 
choices of the masses and coupling.
Scalar fields play a prominent role in many models of cosmology, and their collective excitations could be relevant for transport phenomena in the early universe.
\end{abstract}
%\newpage

\section{Introduction}
Quantum field theory provides the most fundamental description of matter and radiation we know and solves the apparent ``wave particle dualism'' in a consistent way. 
With quantised fields being the fundamental building blocks of nature, 
the elementary excitations of these fields in weakly coupled systems propagate like particles.
Curiously, there are not only elementary particles; in a medium the collective excitations of many elementary quanta often effectively behave as if they were particles themselves.

Often one is not interested in the fate of individual particles, but mostly in transport of energy or charges within 
a system.
Transport phenomena can be studied in a thermodynamic description in terms of a density matrix $\varrho$. 
The propagator in this effective thermodynamic description can have a rather different structure than in vacuum.
This reflects the fact that propagating particles are affected by the medium.
In weakly coupled systems this effect can often be parametrised by interpreting the poles of the propagator as \textit{quasiparticles} with modified properties.
For instance, the dispersion relations (or ``bands'') of electrons in a solid state can be very different from that in vacuum. 
Also the effective charge is screened in a medium. 
In addition to the screened elementary particles there can be new types of quasiparticles that have no analogue in vacuum. These can be interpreted as quantised collective excitations of the background medium. 
For instance, in a solid state the lattice vibrations, phonons, behave like quasiparticles. 
The existence of collective excitations is also well-known from relativistic quantum field theory. 
In gauge theories with coupling $\alpha\ll1$ in thermal equilibrium at high temperature $T$ 
there are fermionic excitations with soft momenta $\p\sim \alpha T$ \cite{Klimov:1981ka,Klimov:1982bv,Weldon:1982bn,Baym:1992eu,Quimbay:1995jn} and ultrasoft momenta $\p\sim \alpha^2 T$ \cite{Satow:2010ia,Miura:2013fxa} which have no analogue in vacuum. These are often referred to as \emph{holes} or \emph{plasminos}. Collective fermionic excitations have also been found in models with Yukawa interactions \cite{Thoma:1994yw,Kitazawa:2006zi,Hidaka:2011rz}.
Also longitudinal gauge bosons appear at finite temperature with a dispersion relation that differs from the transverse components.

In this work we find evidence that collective excitations can also exist in purely scalar field theories.
The existence of collective propagating modes in principle is expected; in particular hydrodynamic modes, such as sound waves, should appear in the spectrum of any field theory.
However, to the best of our knowledge, quasiparticles beyond the hydrodynamic regime have not been described explicitly in the context of purely scalar field theories. 
On one hand their existence can simply be viewed as an interesting property of the field theory.
On the other hand, current experimental evidence \cite{:2012gk,:2012gu} suggests that there is at least one scalar field in nature, the Higgs field. Furthermore, many models of cosmology involve additional scalar fields, such as axions, the inflaton, dilaton, moduli fields or Affleck-Dine fields. Since the universe was exposed to very high temperatures during the early stages of its history, the spectrum of scalar quasiparticles may have affected transport phenomena in the early universe. 

\section{The quasiparticle spectrum in a simple scalar model}
We consider a simple model of two scalar fields described by the Lagrangian
\begin{eqnarray}\label{L}
\mathcal{L}&=& 
\frac{1}{2}\partial_{\mu}\phi\partial^{\mu}\phi
-\frac{1}{2}m_\phi^{2}\phi^{2}
+\frac{1}{2}\partial_{\mu}\chi\partial^{\mu}\chi
-\frac{1}{2}m_\chi^{2}\chi^{2}
-g\phi\chi^2.
\end{eqnarray}
We choose this Lagrangian for illustrative purposes, as it describes the (probably) simplest scalar model in which the additional collective excitations we found appear. We expect that similar behaviour can be found in more realistic models where the structure of the self-energies is similar.
\footnote{Note that the energy functional obtained from (\ref{L}) is not bound from below. For the purpose of illustrating the appearance of collective scalar quasiparticles we will ignore this issue here and consider small excitations around the local minimum at $\phi=\chi=0$, assuming that (\ref{L}) is embedded into a bigger framework that stabilises the ground state.}

\subsection{Quasiparticles in thermal field theory}
Following the approach of \cite{Schwinger:1960qe,Keldysh:1964ud,KBE}
we study the system in terms of real time correlation functions.
This approach has been applied to scalar fields in different situations \cite{Calzetta:1986ey,Calzetta:1986cq,Yokoyama:2004pf,Boyanovsky:2004dj,Anisimov:2008dz,Berges:2008wm,Drewes:2010pf,Berges:2010zv,Garbrecht:2011xw,Hamaguchi:2011jy,Gautier:2012vh,Gautier:2013aoa,Drewes:2012qw,Garbrecht:2011gu,Garbrecht:2013coa,Miyamoto:2013gna} relevant for cosmology.
We use the notation of \cite{Anisimov:2008dz}.
The expectation values or one-point functions $\langle\phi(x)\rangle$ and $\langle\chi(x)\rangle$ play the role of the ``classical field''. 
The average $\langle\ldots\rangle$ is defined in the usual way as $\langle\mathcal{A}\rangle={\rm Tr}(\varrho \mathcal{A})$, where $\varrho$ is the density matrix of the thermodynamic ensemble. It includes the usual quantum average as well as a statistical average over initial conditions.
We will in the following assume that all degrees of freedom are in thermal equilibrium and set $\langle\phi(x)\rangle = \langle\chi(x)\rangle = 0$.
Quasiparticle properties are encoded in the propagator or two-point function. 
We can define two independent two-point functions for $\phi$,
\begin{eqnarray}
\Delta^-(x_1,x_2)&=&i\left(\langle\phi(x_1)\phi(x_2)\rangle-\langle\phi(x_2)\phi(x_1)\rangle\right)\label{deltaminusdefinition}\\
\Delta^+(x_1,x_2)&=&\frac{1}{2}\left(\langle\phi(x_1)\phi(x_2)\rangle+\langle\phi(x_2)\phi(x_1)\rangle\right),
\end{eqnarray}
and analogously for $\chi$.
$\Delta^-$ is called the \textit{spectral function}. It encodes the properties of quasiparticles and is the main quantity of interest in this work.
$\Delta^+$ is called the \textit{statistical propagator} and characterises the occupation numbers of different modes.
Out of thermal equilibrium, $\Delta^-(x_1,x_2)$ and $\Delta^+(x_1,x_2)$ would be two independent functions, and each of them would depend on $x_1$ and $x_2$ individually.
Thermal equilibrium is homogeneous, isotropic and time translation invariant, hence the correlation functions can only depend on the relative coordinate $x_1-x_2$.\footnote{Furthermore, in thermal equilibrium $\Delta^-$ and $\Delta^+$ are not independent, but related by the Kubo-Martin-Schwinger relation, which for their Fourier transforms reads $\Delta^+_\p(p_0)=\frac{1+2f_B(p_0)}{2}\rho_\p(p_0)$. Here $f_B$ is the Bose-Einstein distribution. This is the quantum field theoretical version of the detailed balance relation.} This allows to define the Fourier transform 
\begin{equation}
\rho_\p(p_0)=-i\int d^4(x_1-x_2) \phantom{i} e^{ip_0(t_1-t_2)} e^{-i\p(\textbf{x}_1-\textbf{x}_2)} \Delta^-(x_1-x_2)\label{spectraldensitydefinition}
.\end{equation}
It can be expressed as \cite{Anisimov:2008dz}
\begin{eqnarray}\label{spectralfunction2}
\rho_{\p}(p_0)={-2{\rm Im}\Pi^R_{\p}(p_0)+2p_0\epsilon\over 
(p_0^2-m^2-\p^2-{\rm Re}\Pi^R_{\p}(p_0))^2+({\rm Im}\Pi^R_{\p}(p_0)+p_0\epsilon)^2}. 
\end{eqnarray}
Here $\Pi^R_\p(p_0)$ is the Fourier transform of the usual retarded self-energy, in this case
\begin{equation}
\Pi^R_\phi(x_1,x_2)=g^2\theta(t_1-t_2)\Big(\chi(x_1)\chi(x_1)\chi(x_2)\chi(x_2)
-\chi(x_2)\chi(x_2)\chi(x_1)\chi(x_1)\Big)
,\end{equation}
and analogous for $\chi$.
In (\ref{spectralfunction2}) we have not specified whether we refer to $\phi$ or $\chi$; both spectral densities formally have the same shape except for the replacement $m\rightarrow m_\phi$ or $m\rightarrow m_\chi$ and the insertion of the corresponding self-energy.
The pole structure of $\rho_\p(p_0)$ in the complex $p_0$ plane determines the spectrum of quasiparticles.
In vacuum there would be only one pole for positive $p_0$ at $p_0=\omega_\p\equiv(\p^2+m^2)^{1/2}$, where $m$ is the renormalised mass.
At $T>0$ there can be several poles, which we will label by an index $^i$.
We refer to the pole that converges to $\omega_\p$ in the limit $T\rightarrow 0$ as the \emph{screened one-particle state}, and to all other poles as purely collective excitations. 
 
$\Pi^R_\p(p_0)$ can be expressed as the sum of a vacuum contribution and a temperature dependent medium correction.
The real part of the vacuum contribution contains the usual UV divergence that also appears in vacuum, the temperature dependent part is UV-finite. It is common to impose renormalisation conditions at $T=0$ to absorb the divergence and define the physical mass \cite{Boyanovsky:2004dj,Anisimov:2008dz}. 
We will in the following simply interpret $m_\phi$ and $m_\chi$ as physical masses in vacuum after renormalisation and ${\rm Re}\Pi^R_\p(p_0)$ as the remaining finite piece.\footnote{Formally we should use different symbols for the mass parameter appearing in (\ref{L}) and full self-energy before renormalisation on one hand and the physical mass and finite part of ${\rm Re}\Pi^R$ on the other. However, the former do not appear anywhere in the following calculation.} 
Let $\hat{\Omega}_\p^i$ be a pole of $\rho_\p(p_0)$ with $\Omega_\p^i\equiv{\rm Re}\hat{\Omega}_\p^i$ and $\Gamma_\p^i\equiv2{\rm Im}\hat{\Omega}_\p^i$.
$\Omega_\p^i$ and $\Gamma_\p^i$ are temperature dependent because $\Pi^R_\p(\omega)$ depends on $T$.
In weakly coupled theories one usually observes the hierarchy
\begin{equation}\label{quasiparticle}
\Gamma^i_\p\ll\Omega^i_\p .
\end{equation}
Due to (\ref{quasiparticle}) we can interpret $\Omega_\p^i$ as a quasiparticle\footnote{We refer to any pole of a propagator that fulfils (\ref{quasiparticle}) as quasiparticle, may it be a screened one-particle state or a collective excitation, and regardless of its spin.} dispersion relation (or ``thermal mass shell'') and $\Gamma_\p^i$ as its thermal width (or damping rate). 
Near poles that fulfil (\ref{quasiparticle}) the spectral density can be approximated by 
\begin{equation}\label{BW}
\rho_{\textbf{p}}^{{\rm BW}}(p_0)\big|_{p_0\simeq \Omega_{\textbf{p}}^{i}} \simeq \sum_i 2\mathcal{Z}_{\p}^{i}\frac{p_0\Gamma_{\textbf{p}}^{i}}{\big(p_0^2-(\Omega_{\textbf{p}}^{i} )^2\big)^2+\big(p_0\Gamma_{\textbf{p}}^{i}\big)^2} + \rho_{ \textbf{p}}^{\rm cont}(p_0)
\end{equation}
Here the residue and width are given by 
\begin{equation}\label{ZandGamma}
\mathcal{Z}_\p^{i}=\left[1-\frac{1}{2\Omega_\p^{i}}\frac{\partial {\rm Re}\Pi^R_\p(p_0)}{\partial p_0}\right]^{-1}_{p_0=\Omega_\p^{i}} \ , \ \Gamma_\p^{i}=-\mathcal{Z}_\p^{i}\frac{{\rm Im}\Pi^R_\p(\Omega_\p^{i})}{2\Omega_\p^{i}}
.\end{equation}
In the zero-width limit the it reads
\begin{equation}\label{zerowidth}
\rho_{\textbf{p}}^{{\rm 0}}(p_0)=\sum_i\mathcal{Z}_{\p}^{i}
2\pi {\rm sign}(p_0)\delta\big(p_0^2-(\Omega_\p^{i})^2\big)
+ \rho_{ \textbf{p}}^{\rm cont}(p_0),
\end{equation}
which can be compared to the free spectral density
\begin{equation}\label{rhofree}
\rho_{\textbf{p}}^{{\rm free}}(p_0)=
2\pi {\rm sign}(p_0)\delta(p_0^2-\omega_\p^2).
\end{equation}
The dispersion relation in (\ref{zerowidth}) is essentially fixed by ${\rm Re}\Pi^R_{\p}(p_0)$ via the condition 
\begin{equation}\label{dispersionrelation}
p_0^2-\p^2-m^2-{\rm Re}\Pi^R_{\p}(p_0)=0. 
\end{equation}
For this reason the real and imaginary part of the retarded self-energy are often referred to as the 
dispersive self-energy and dissipative self-energy, respectively.

The dispersion relations $\Omega_\p^i$ can have a complicated $\p$-dependence. 
In limited momentum regimes they can often be approximated by momentum independent ``thermal masses''. %For the zero-mode one can define the {\it plasma frequency}, i.e. the solution to (\ref{dispersionrelation}) for $\p=0$.
For hard modes $\p\sim T$ it is common to define the {\it asymptotic mass} $M$, which depends on $T$ but not on $\p$, by fitting the approximation $(\p^2+M^2)^{1/2}$ to the full dispersion relation in the regime $\p\gtrsim T$. This approximation is commonly used in transport equations because most particles in a plasma in thermal equilibrium have momenta $\p\sim T$. 
In this work we are interested in collective excitations. These usually appear in the momentum regime $\p\ll T$, where the energy related to inter-particle forces can be comparable to their kinetic energy or larger.
Therefore we cannot use this approximation.
\begin{figure}
  \centering
    \includegraphics[width=12cm]{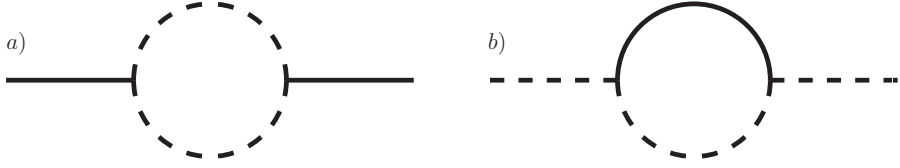}
    \caption{Diagrams contributing to the self-energies for $\phi$, $a)$, and $\chi$, $b)$, at on-loop order. Solid lines represent $\phi$-propagators, dashed lines $\chi$-propagators.\label{diagram}}
\end{figure}
\subsection{The novel scalar plasmons}
For the Lagrangian (\ref{L}) the leading order contribution to the $\phi$-self-energy comes from the diagram shown in figure \ref{diagram}a). Using finite temperature Feynman rules \cite{LeB}, the imaginary part of this diagram can be calculated from
\begin{equation}\label{trilinear}
{\rm Im}\Pi^R_{\phi\p}(p_0)=-\frac{g^2}{4}\int\frac{d^4k}{(2\pi)^4}\left(1+f_B(k_0)+f_B(p_0-k_0)\right)\rho_{\chi \textbf{k}}(k_0)\rho_{\chi \p-\textbf{k}}(p_0-k_0),
\end{equation}
where the subscript indicates the self-energy or spectral density of which field we mean.
At one-loop level the integral (\ref{trilinear}) is to be evaluated with free spectral densities (\ref{rhofree}). This corresponds to using free thermal propagators in the loop.
The result is 
\begin{comment}
\begin{eqnarray}
-{\rm Im}\Pi_{\phi\textbf{p}}^R(p_0)&=&%\theta(p_0)
\frac{g^2}{32\pi}
\Bigg[\left(
-\sqrt{1-\frac{(2m_\chi)^2}{p^2}}
+2\frac{T}{|\textbf{p}|}\log\bigg[
\frac{f_B(\omega_\phi^-)}{f_B(\omega_\phi^+)}
\bigg]
\right)\theta\big(p^2-(2m_\chi)^2\big)\nonumber\\
&&\phantom{\theta(p_0)\frac{g^2}{32\pi}\sqrt{1-\frac{(2m_\chi)^2}{p^2}}}
+\frac{2}{|\textbf{p}|} 
\left(p_0+T{\rm log}\bigg[
\frac{f_B(\omega_\phi^-)}{f_B(-\omega_\phi^+)}
\bigg]
\right)\theta\big(-p^2\big)
\Bigg]
\end{eqnarray}
\end{comment}
\begin{eqnarray}\label{xiIntSelfEn}
\lefteqn{ - \text{Im} \Pi^R_{\phi,\mathbf{p}}=\frac{g^2}{16 \pi |\mathbf{p}|} 
\Bigg[- \theta(-p^2) \Bigg( p_0 + T \log\bigg[\frac{f_B(p_0-\omega_\phi^+)f_B(-\omega_\phi^-)}{f_B(-\omega_\phi^+)f_B(p_0-\omega_\phi^-)}\bigg]\Bigg)
}\\
&+&
\theta\big(p^2 - (2m_\chi)^2\big) 
T \log\bigg[\frac{f_B(p_0-\omega_\phi^+)f_B(-\omega_\phi^-)}{f_B(-\omega_\phi^+)f_B(p_0-\omega_\phi^-)}\bigg]
 \Bigg]\nonumber
\end{eqnarray}
with the Bose-Einstein distribution $f_B(p_0)\equiv(e^{p_0/T}-1)^{-1}$ and
\begin{eqnarray}
\omega_\phi^\pm=\frac{p_0}{2}\pm{\rm sign}(p^2)\frac{|\textbf{p}|}{2}\sqrt{
1-\frac{(2m_\chi)^2}{p^2}}
\end{eqnarray}
The real and imaginary part of the self-energy are related by the Kramers-Kronig relations, which allow to rewrite (\ref{dispersionrelation}) as 
\begin{equation}
p_0^2-\textbf{p}^2-m_\phi^2-\mathcal{P}\int\frac{d \omega}{\pi}\frac{{\rm Im}\Pi^R_{\phi\p}(\omega)}{\omega-p_0}=0.\label{disprelfinder}
\end{equation}
In (\ref{disprelfinder}) we neglect the $T=0$ part of (\ref{xiIntSelfEn}), as this contribution has been absorbed into $m_\phi$ already. 
\begin{figure}
  \centering
    \includegraphics[width=12cm]{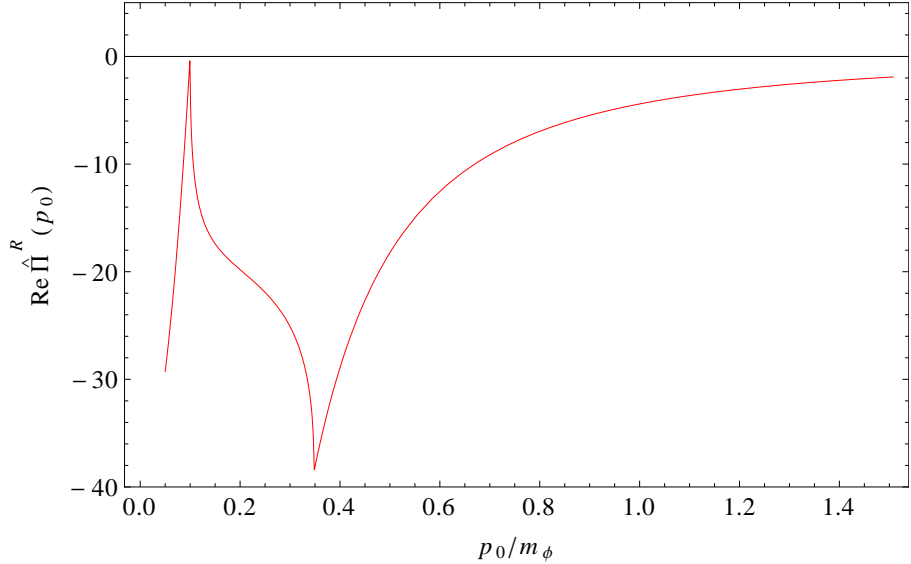}
    \caption{The real part of the  retarded $\phi$-self-energy ${\rm Re}\hat{\Pi}^R_{\phi\p}(p_0)\equiv{\rm Re}\Pi^R_{\phi\p}(p_0)/g^2$ for 
$m_\chi=m_\phi/6$, 
$T=333 m_\phi$ and 
$|\p|=m_\phi/10$ as a function of $p_0$. 
The function is positive for $p_0 \gg m_\phi$.
\label{RePiPlot}}
\end{figure}
\begin{figure}
  \centering
    \includegraphics[width=12cm]{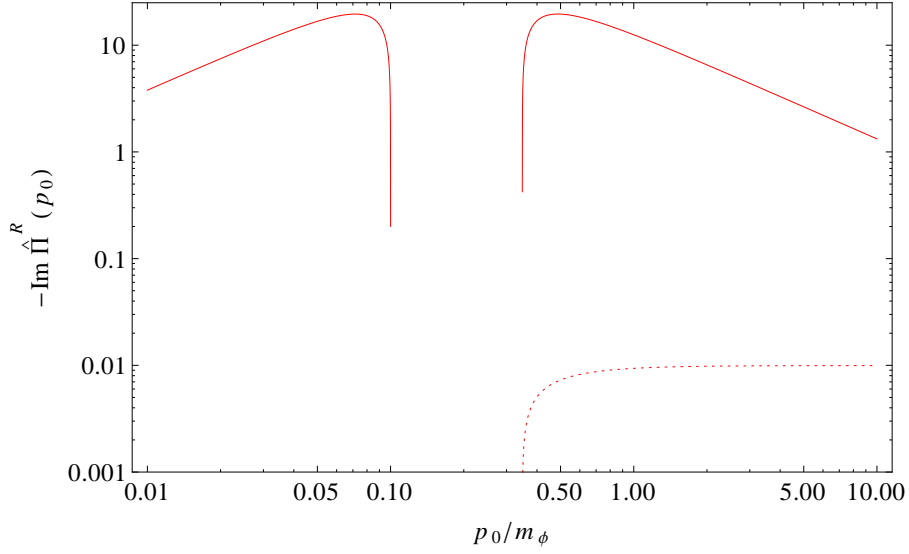}
    \caption{The imaginary part of the  retarded $\phi$-self-energy ${\rm Im}\hat{\Pi}^R_{\phi\p}(p_0)\equiv{\rm Im}\Pi^R_{\phi\p}(p_0)/g^2$ for 
$m_\chi=m_\phi/6$, $T=333m_\phi$ and $|\p|=m_\phi/10$ as a function of $p_0$. The dotted line is the $T=0$ contribution.
\label{ImPiPlot}}
\end{figure}

\begin{figure}
  \centering
    \includegraphics[width=12cm]{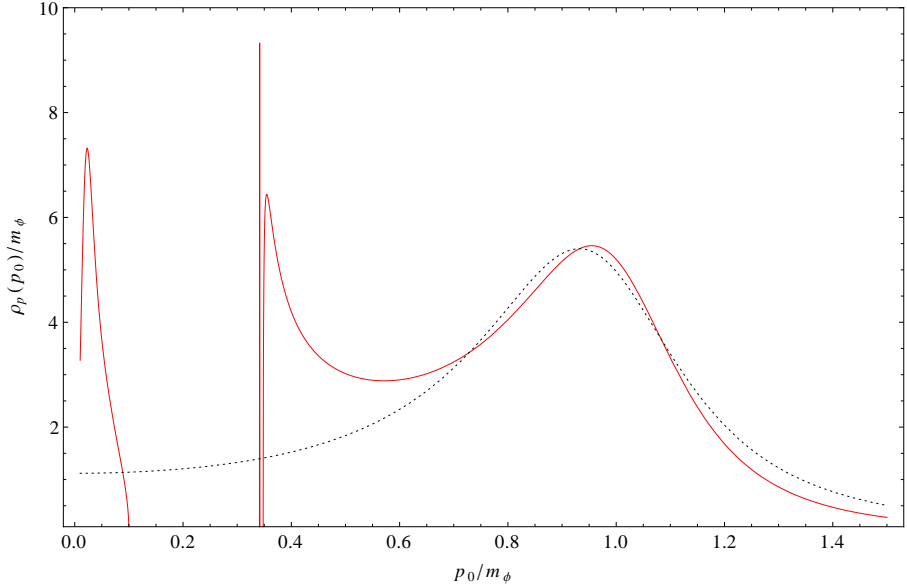}
    \caption{Solid red line: The spectral density 
$\rho_{\phi\p}(p_0)$ from (\ref{spectralfunction2}) with 
$g=m_\chi=m_\phi/6$, $T=333m_\phi$ and $|\p|=m_\phi/10$ as a function of $p_0$. The isolated feature near $p_0\simeq 0.34 m_\phi$ represents a $\delta$-function, which we interpret as a novel scalar plasmon of purely collective origin, the \emph{luon}. 
Another interesting feature is the continuous contribution for $p^2<0$, which disappears in the limit $T\rightarrow 0$. 
Dotted black line: The Breit-Wigner approximation (\ref{BW}) in the vicinity of the 
screened particle pole.
\label{rhoplot}}
\end{figure}

We now choose a set of parameters $g=m_\chi=m_\phi/6$, $T=333m_\phi$ and $|\p|=m_\phi/10$. 
The real and imaginary part of the self-energy for this choice are shown in figure \ref{RePiPlot} and \ref{ImPiPlot}. 
The most prominent features in ${\rm Re}\Pi^R_{\phi\p}(p_0)$  are two spikes, which appear near $p_0^2=\p^2+(2m_\chi)^2$ because the zero in the denominator of the principal value term in (\ref{disprelfinder}) passes the kinematic thresholds of (\ref{xiIntSelfEn}) in the numerator. Another remarkable feature is that the finite temperature correction is negative in that $p_0$-region. 
This leads to a negative thermal mass correction, and it is precisely the reason why we find more than one solution to (\ref{disprelfinder}). 
The negative mass shift seems unusual from a particle physics viewpoint, where one is used to positive ``thermal masses''. Note, however, that a frequency $\Omega_\p$ that is smaller than in vacuum for a given $\p$ in optics corresponds to a index of refraction greater than one, i.e. the normal behaviour. 
\begin{figure}
  \centering
    \includegraphics[width=12cm]{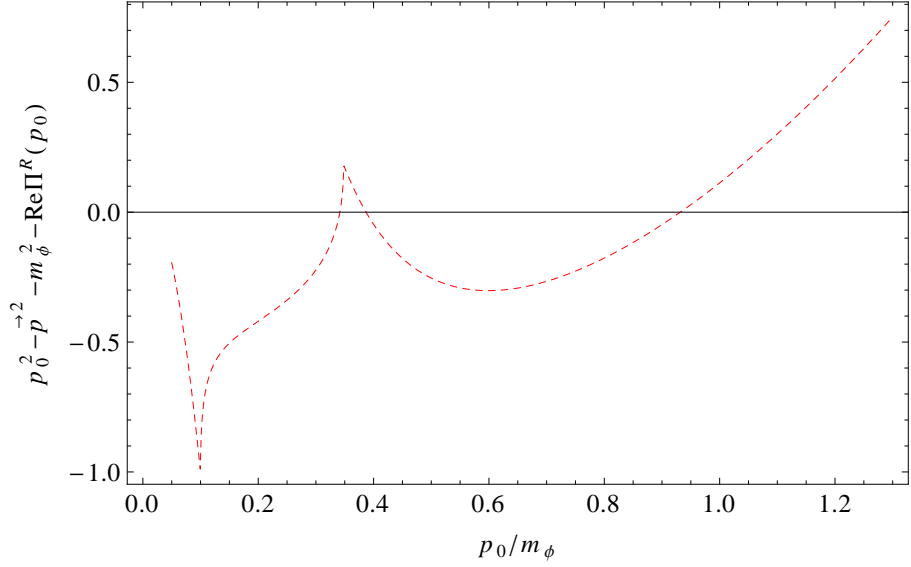}
    \caption{The left hand side of (\ref{disprelfinder}) for $g=m_\chi=m_\phi/6$, $T=333m_\phi$ and $|\p|=m_\phi/10$ as a function of $p_0$. 
\label{mustbezero}}
\end{figure}
The left hand side of (\ref{disprelfinder}) is plotted in figure \ref{mustbezero}.
There are three solutions to (\ref{disprelfinder}), which we label by $\Omega_\p^a$, $\Omega_\p^b$ and $\Omega_\p^c$. One of them is at $p_0\simeq \omega_\p$ and can be interpreted as dressed one-particle state. The resulting spectral density $\rho_{\phi,\p}(p_0)$ is shown in figure \ref{rhoplot}.

The first peak from the right lies at $\Omega_\p^a\simeq\omega_\p$. It is the dressed one-particle state. 
It is interesting that this pole lies so close to the vacuum mass shell. This implies that thermal mass corrections are negligible for low momentum modes in the model described by (\ref{L}), which has already been observed in \cite{Drewes:2013iaa}. This is also confirmed by the analysis in appendix \ref{DimensionalReduction}.
Naively one might have expected that the thermal correction dominates over the vacuum mass for $gT\gg m_\phi^2$.
In spite of the small thermal mass corrections, thermal effects strongly dominate the width: $\Gamma_\p^a$ is about $\sim 1400$ times bigger than the vacuum decay width,\footnote{In figure \ref{ImPiPlot} it can be seen that thermal effects strongly dominate ${\rm Im}\Pi^R_{\phi,\p}(p_0)$.} leading to a significant broadening of the peak. 
With $\Gamma_\p^a/\Omega_\p^a\simeq 0.4$ the screened one-particle state is a short lived resonance, but the approximation of $\rho_\p(p_0)$ by the Breit-Wigner function (\ref{BW}) near the peak works at reasonable accuracy, see figure \ref{rhoplot}.

The remaining two poles at $\Omega_\p^b\simeq 0.39 m_\phi$ and $\Omega_\p^c\simeq 0.34 m_\phi$ are due to the negative spike in ${\rm Re}\Pi^R_\p(p_0)$.
They lie very close to each other; their separation is considerably smaller than $\Gamma_\p^a$. They do, however, not merge into a single resonance because 
$\Omega_\p^c$ has vanishing width in the one-loop approximation (\ref{xiIntSelfEn}). This is because (\ref{xiIntSelfEn}) vanishes below the two-particle threshold, as the only processes contributing to ${\rm Im}\Pi^R_{\phi,\p}(p_0)$ at one-loop level on-shell are decays and inverse decays $\phi\leftrightarrow \chi\chi$. Therefore $\Omega_\p^c$ is associated with a stable quasiparticle in this approximation. 
We refer to this novel, purely collective \emph{l}ow moment\emph{u}m scalar plasm\emph{on} as \emph{luon}.

The solution $\Omega_\p^b$, on the other hand, is broadened and absorbed by $\rho_{\phi,\textbf{p}}^{\rm cont}(p_0)$ near the threshold, where ${\rm Im}\Pi^R_{\phi,\p}(p_0)$ is relatively large.
This implies that there is no propagating quasiparticle associated with $\Omega_\p^b$, as the lifetime of this resonance would be so short that its mean free path in the plasma is shorter than its de Broglie wavelength. 
The Breit-Wigner approximation is not valid near $\Omega_\p^b$ because (\ref{xiIntSelfEn}) is a steep function in the regime near the two-particle threshold. 
Hence, we keep the full $p_0$-dependence of $\Pi^R_{\phi,\p}(p_0)$ in (\ref{spectralfunction2}) in all plots.
This is also obvious from the fact that the quasiparticle associated with a similar energy $\Omega_\p^c$ in the very same approximation (\ref{xiIntSelfEn}) has infinite lifetime. 
It shows that the spectrum of quasiparticles is very sensitive to the threshold behaviour of ${\rm Im}\Pi^R_\p(p_0)$, which may be affected by higher order corrections.

\section{Discussion}
We have studied the spectrum of quasiparticles in a hot scalar field theory.
We calculated the spectral density $\rho_{\phi\p}(p_0)$ at one-loop level. 
For our choice of parameters we find that $\rho_{\phi\p}(p_0)$  has three poles for some momenta in the regime $\p< m_\phi$. One of them can clearly be identified with the dressed one-particle state. It shows significant thermal broadening, but only a negligible thermal mass shift. It can be described by a Breit-Wigner approximation. 
Of the remaining two poles, one can be interpreted as a new quasiparticle while the other one is broad and cannot be interpreted as a quasiparticle.
Both have very similar energies and lie in the threshold region. Therefore our results are rather sensitive to loop corrections in this region. 
\subsection{Consistency of our result}
As a first cross-check for the validity of our result, we evaluate the sum rule
\begin{equation}\label{sumrule}
\int\frac{dp_0}{2\pi}p_0\rho_\p(p_0)=1,
\end{equation}
which directly follows from the commutation relations $[\phi(x_1),\dot{\phi}(x_2)]|_{t_1=t_2}=i\delta(\textbf{x}_1-\textbf{x}_2)$ for a scalar field.
Numerical evaluation of (\ref{sumrule}) with (\ref{spectralfunction2}) yields $0.996$ for our parameter choice, i.e. good agreement.
In order to estimate the importance of the different features in $\rho_{\phi\p}(p_0)$ for transport, it is instructive to evaluate the contribution to (\ref{sumrule}) from different $p_0$-regions separately.
The region $p^2<0$ contributes less than $1\%$. 
The luon at $\Omega_\p^c$ contributes about $6\%$. 
This suggests that the collective excitations generally do not contribute much to phase space integrals. 
The exchange of luons will typically affect transport in the plasma even less than in the case of fermionic holes in gauge theories:
both types of collective excitations only exist for small momenta $\p\ll T$ (while most particles in a relativistic plasma have momenta $\p \sim T$), but the fermionic holes have residues of order one for soft momenta. 
The region around $\Omega_\p^b$ contributes about $8\%$ to the sum rule (\ref{sumrule}). The main contribution of $72\%$ comes from the region around the screened one-particle pole at $\Omega_\p^a$. This suggests that transport in the low momentum region predominantly happens via the exchange of these short lived resonances. Finally, there is a contribution of $13\%$ from the continuum contribution 
at $p_0>\Omega_\p^a$.

In spite of their small residue luons could make a relevant contribution to transport when the processes involving the dressed particles 
are kinematically forbidden. 
A situation of this kind has been studied in detail in \cite{Drewes:2013iaa} for the relaxation of the zero-mode of a massive scalar field in a plasma of scalars and fermions with gauge interactions at temperatures larger than the scalar's mass. The scalar may be identified with the inflaton, an axion, moduli-field or an order parameter during a phase transition.
The relaxation rate is sensitive to the quasiparticle spectrum at momenta $\k\ll T$ because the momenta of the decay products are of the order of the scalar mass, and not of the order of the temperature. 
In some temperature regimes the processes involving dressed particles can be kinematically blocked by large thermal masses of the decay products. 
It was found in \cite{Drewes:2013iaa} that in this situation it is crucial to take into account the full quasiparticle spectrum at low momentum.\footnote{
Also in scalar QED the dissipation rate is sensitive to the spectrum in the infrared. 
At one-loop level it is formally of order $\alpha^2$ \cite{Miyamoto:2013gna} due to processes $\phi\leftrightarrow\phi\gamma$: While in vacuum the decay $\phi\rightarrow\phi\gamma$ is kinematically forbidden, it formally leads to a finite dissipation rate at $T\neq 0$ because the vanishing phase space volume is compensated by the infinite occupation number for photons with energy zero. This unphysical behaviour arises because the thermal photon mass has been neglected, which regularises the infinite occupation number leads to the kinematic blocking. 
Once the thermal photon mass is included the on-shell decay of a scalar particle into itself and a photon cannot contribute to the dissipation any more.
If collective scalar plasmons are present in the spectrum, then the decay of a screened scalar particle into a photon and such a plasmon could still give a non-zero contribute (along with different scattering processes obtained from other cuts through the self-energies \cite{Drewes:2010pf,Garbrecht:2010sz,Anisimov:2010dk,Drewes:2013iaa,Salvio:2013iaa}).
}
While this raises hope that there may be physical systems in which the exchange of luons is relevant, it also forces us to question whether the use of 
the one loop result (\ref{xiIntSelfEn}) is justified.\footnote{The parameters used in the examples given in \cite{Drewes:2013iaa} were deliberately chosen to avoid the issues we discuss in the following. }
The new scalar plasmons lie in the threshold region, where the shape of the self-energy and spectral density are usually sensitive to corrections of higher order in the loop expansion. Some of these corrections can be taken into account by using full $\chi$-propagators in the loop in figure \ref{diagram}a).\footnote{In addition, there may be vertex corrections due to ``ladder diagrams'' of the same order, which we do not discuss here. 
We assume that they simply lead to a change in the effective coupling constant or are of higher order. For Yukawa type vertices it has been argued that this assumption is applicable to diagrams of this topology \cite{Thoma:1994yw}, but for gauge interactions it is not \cite{Anisimov:2010gy}.
}
This is expected to have two different effects.

On one hand the full $\chi$-propagators include a finite width due to the imaginary part of the thermal $\chi$-self-energies. The expression (\ref{xiIntSelfEn}) for ${\rm Im}\Pi^R_{\phi\p}(p_0)$
obtained using free thermal $\chi$-propagators  vanishes below the two-particle threshold $p_0^2=\p^2+(2m_\phi)^2$, see figure \ref{ImPiPlot}. It is clear that  ${\rm Im}\Pi^R_{\phi\p}(p_0)$ is non-vanishing along the entire $p_0$ axis once contributions of higher order in the loop expansion are taken into account \cite{Drewes:2010pf,Drewes:2013iaa}, which can be interpreted as damping by scatterings and lead to a $\Gamma_\p^c\neq 0$. 
This width smears out the kinematic thresholds \cite{Drewes:2010pf,Drewes:2013iaa} and leads to a non-vanishing ${\rm Im}\Pi^R_{\phi\p}(p_0)$ at  $p_0^2<\p^2+(2m_\phi)^2$.
Due to the very small splitting $\Omega_\p^b-\Omega_\p^c$ it seems possible that the peaks in $\rho_\p(p_0)$ at $\Omega_\p^b$ and $\Omega_\p^c$  merge and effectively form a single resonance when these processes are taken into account. Whether or not this resonance is sufficiently long-lived to be interpreted as a propagating quasiparticle depends on the size of the corrections.
Another effect of smoothing out the kinematic thresholds is that the spike in ${\rm Re}\Pi^R_{\phi,\p}(p_0)$ is less sharp. This is because the spike appears where the pole in the denominator of the integral in (\ref{disprelfinder}) passes the kinematic threshold.
Also this tends to have the effect of broadening the resonances. 
The size of $\Gamma_\p^c$ cannot be determined without a proper calculation. 
It is known that in some cases the effects of multiple scatterings entirely overcome the suppression of ${\rm Im}\Pi^R_{\phi,\p}(p_0)$ in the region that is kinematically forbidden at one-loop level \cite{Anisimov:2010gy}. This would possibly eliminate the luon from the quasiparticle spectrum. However, there are also situations in which the suppression remains when higher order corrections are taken into account \cite{Drewes:2013iaa}.

The other important effect of using dressed $\chi$-propagators are the modifications to the $\chi$-dispersion relations due to the real part of the $\chi$-self-energy.
Thermal mass corrections can shift the two-particle threshold in ${\rm Im}\Pi^R_{\phi,\p}(p_0)$, and hence the position of the spike in ${\rm Re}\Pi^R_{\phi,\p}(p_0)$, to larger values of $p_0$. This is crucial as ${\rm Re}\Pi^R_{\phi,\p}(p_0)$  can compete with $p_0^2-\textbf{p}^2-m_\phi^2$ in (\ref{disprelfinder}) only because of the spike, and it can do so only if the spike lies in the region $p_0\lesssim m_\phi$, where $p_0^2-\textbf{p}^2-m_\phi^2$ is small. 
If thermal masses would shift the spike to the region $p_0\gg m_\phi$, then ${\rm Re}\Pi^R_{\phi,\p}(p_0) < p_0^2-\textbf{p}^2-m_\phi^2$ for all $p_0$, and there is only one solution to (\ref{disprelfinder}). 
The analysis in the following section and appendix \ref{DimensionalReduction} suggests that this is not the case for our choice of parameters.

\subsection{The $\chi$ propagator at one loop level}
\begin{figure}
  \centering
    \includegraphics[width=12cm]{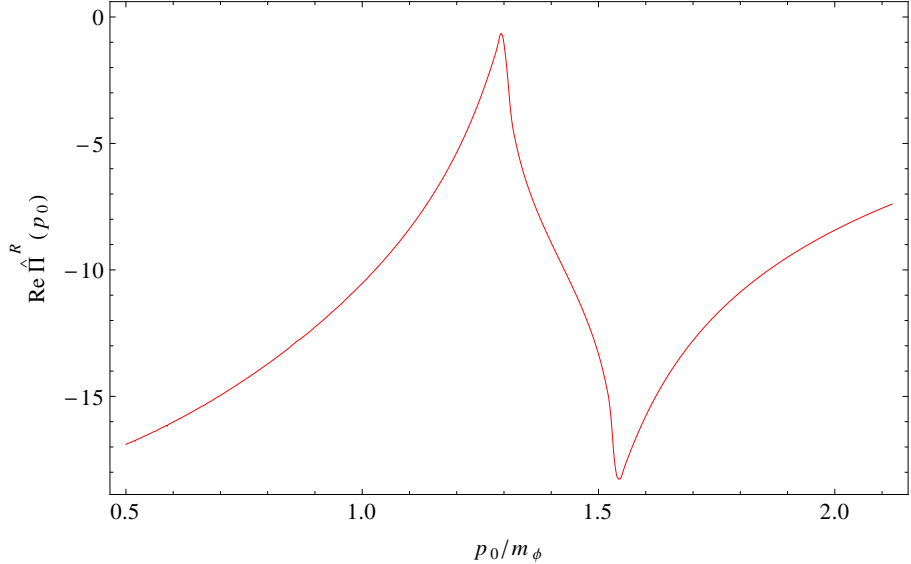}
    \caption{The real part of the  retarded $\chi$-self-energy ${\rm Re}\hat{\Uppi}^R_{\chi\p}(p_0)\equiv{\rm Re}\Uppi^R_{\chi\p}(p_0)/g^2$ for $g=m_\chi=m_\phi/6$, $T=333m_\phi$ and $|\p|=m_\phi$ as a function of $p_0$. 
\label{RePiPlotchi}}
\end{figure}

To estimate the thermal corrections to $\chi$-properties we evaluate the $\chi$-self energy $\Uppi^R_{\chi\p}(p_0)$ in figure \ref{diagram}b) with free thermal propagators in the loop, i.e. using free spectral densities (\ref{rhofree}).
The imaginary part is given by 
\begin{equation}\label{trilinear2}
{\rm Im}\Uppi^R_{\chi,\k}(k_0)=-\frac{g^2}{2}\int\frac{d^4q}{(2\pi)^4}\left(1+f_B(q_0)+f_B(k_0-q_0)\right)\rho_{\phi \textbf{q}}(q_0)\rho_{\chi \k-\textbf{q}}(k_0-q_0)
\end{equation}
and can again be computed analytically,
\begin{comment}
\begin{eqnarray}\label{Uppi}
\lefteqn{
-\im\Uppi^{R}_{\chi,\k}(k_0) = \frac{g^2}{16\pi|\k|}
\Bigg[
\Big(\omega_\chi^--\omega_\chi^+\Big) \theta\big(k^2-(m_\chi+m_\phi)^2\big)
}
\nonumber\\
&+&T {\rm log}\bigg[
\frac{f_B(k_0-\omega_\chi^+)f_B(\omega_\chi^-)}{f_B(\omega_\chi^+)f_B(k_0-\omega_\chi^-)}
\bigg]\theta\big(k^2-(m_\chi+m_\phi)^2\big)\nonumber\\
&+&
T {\rm log}\bigg[
\frac{f_B(k_0-\omega_\chi^+)f_B(-\omega_\chi^-)}{f_B(-\omega_\chi^+)f_B(k_0-\omega_\chi^-)}
\bigg]\theta\big(-k^2+(m_\chi-m_\phi)^2\big)\theta(k^2)\nonumber\\
&+& 
T {\rm log}\bigg[
\frac{f_B(k_0-\omega_\chi^+)f_B(\omega_\chi^-)}{f_B(-\omega_\chi^+)f_B(\omega_\chi^--k_0)}
\bigg]\theta\big(-k^2+(m_\chi-m_\phi)^2\big)\theta(-k^2)
\nonumber\\
&+&2k_0\theta\big(-k^2+(m_\chi-m_\phi)^2\big)\theta(-k^2)\Bigg]
\end{eqnarray}
with
\begin{eqnarray}
\omega_\chi^\pm = \frac{1}{2k^2}\bigg[
k_0(m_\chi^2-m_\phi^2+k^2)\pm
|\k|\sqrt{
m_\chi^4+(m_\phi^2-k^2)^2-2m_\chi^2(m_\phi^2+k^2)
}
\bigg]
\end{eqnarray}
Here the first line is a zero-temperature contribution due to the decay 
process $\chi\rightarrow\chi\phi$, which are of course kinematically forbidden on-shell. The second line contains a the finite-temperature contribution from this process and its inverse from induced transitions.
The third and fourth line include the finite-temperature contributions from processes $\phi\leftrightarrow\chi\chi$.
\end{comment}
\begin{eqnarray}\label{Uppi}
\lefteqn{ - \text{Im} \Pi^R_{\chi,\mathbf{k}}=\frac{g^2}{16 \pi |\mathbf{k}|} 
\Bigg[- \theta(-k^2) \, k_0}\\
&+&\Big( \theta\big(k^2 - (m_\phi+m_\chi)^2\big) - \theta\big(- k^2 + (m_\phi-m_\chi)^2\big)\Big) 
T \log\bigg[\frac{f_B(k_0-\omega_\chi^+)f_B(-\omega_\chi^-)}{f_B(-\omega_\chi^+)f_B(k_0-\omega_\chi^-)}\bigg]
 \Bigg]\nonumber
\end{eqnarray}
with 
\begin{equation}
\omega_{\chi}^{\pm}=\frac{1}{2 k^2}\left[k_0(m_{\phi}^2 - m_{\chi}^2+k^2)\pm |\mathbf{k}|
\sqrt{m_{\chi}^4+(m_{\phi}^2-k^2)^2-2 m_{\chi}^2 (m_{\phi}^2+k^2)} \right]
\end{equation}
Due to the different processes, there are two thresholds.
This leads to the appearance of a double-spike in the real part in figure \ref{RePiPlotchi}, which we again calculate using the Kramers-Kronig relations. Since both thresholds involve the heavier mass $m_\phi$, the spikes lie in a region where they cannot compete with $k_0^2-\k^2-m_\chi^2$, hence there is no collective excitation. We now use this result to discuss the spectrum of $\chi$-quasiparticles.

\begin{figure}
  \centering
    \includegraphics[width=12cm]{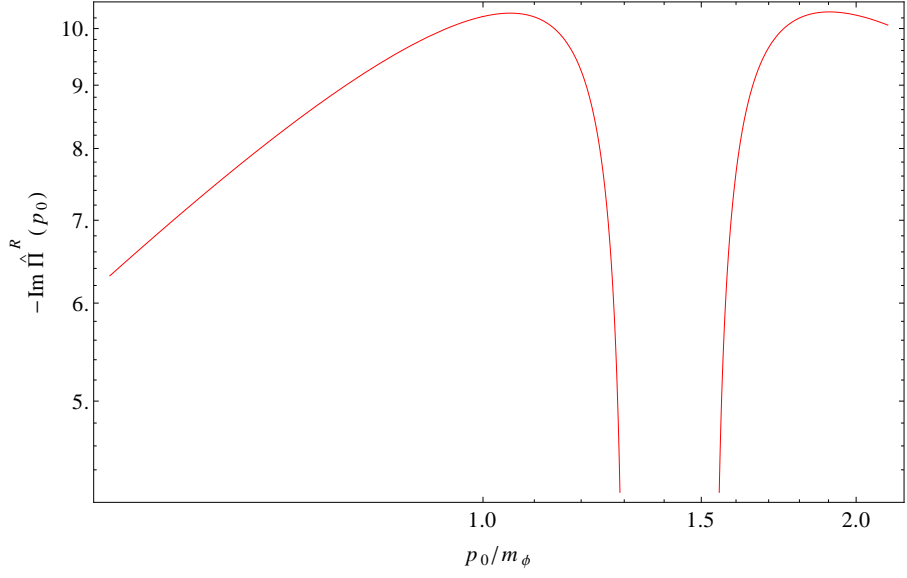}
    \caption{The imaginary part of the  retarded $\chi$-self-energy ${\rm Im}\hat{\Uppi}^R_{\chi\p}(p_0)\equiv{\rm Im}\Uppi^R_{\chi\p}(p_0)/g^2$ for 
$g=m_\chi=m_\phi/6$, $T=333m_\phi$ and $|\p|=m_\phi$ as a function of $p_0$. %The dotted line i the $T=0$ contribution.
\label{ImPiPlotchi}}
\end{figure}

\begin{figure}
  \centering
    \includegraphics[width=12cm]{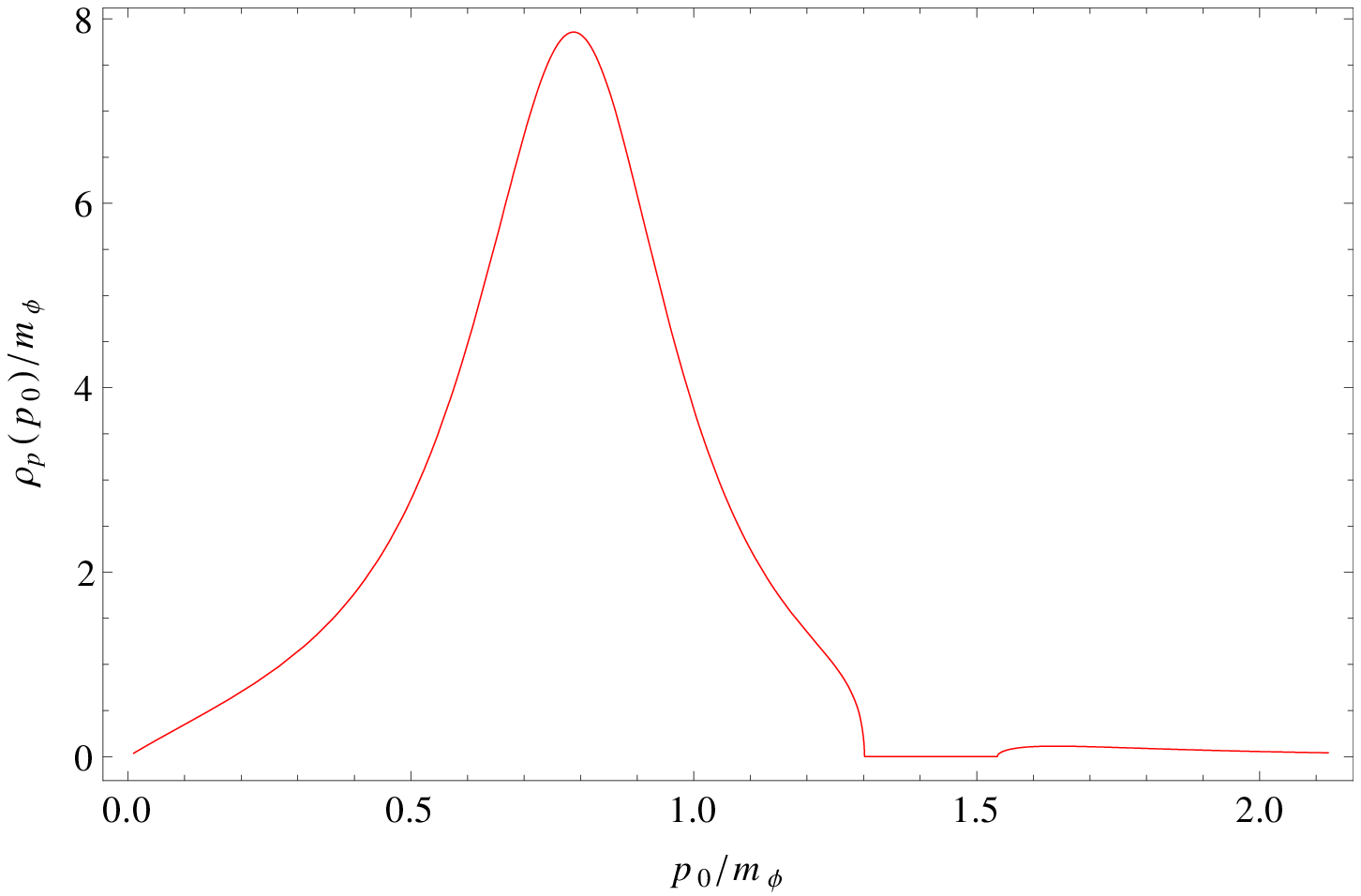}
    \caption{The spectral density $\rho_{\chi\p}(p_0)$ with $g=m_\chi=m_\phi/6$, $T=333m_\phi$ and $|\p|=m_\phi$ as a function of $p_0$. 
\label{rhoplotchi}}
\end{figure}

\begin{figure}
  \centering
    \includegraphics[width=12cm]{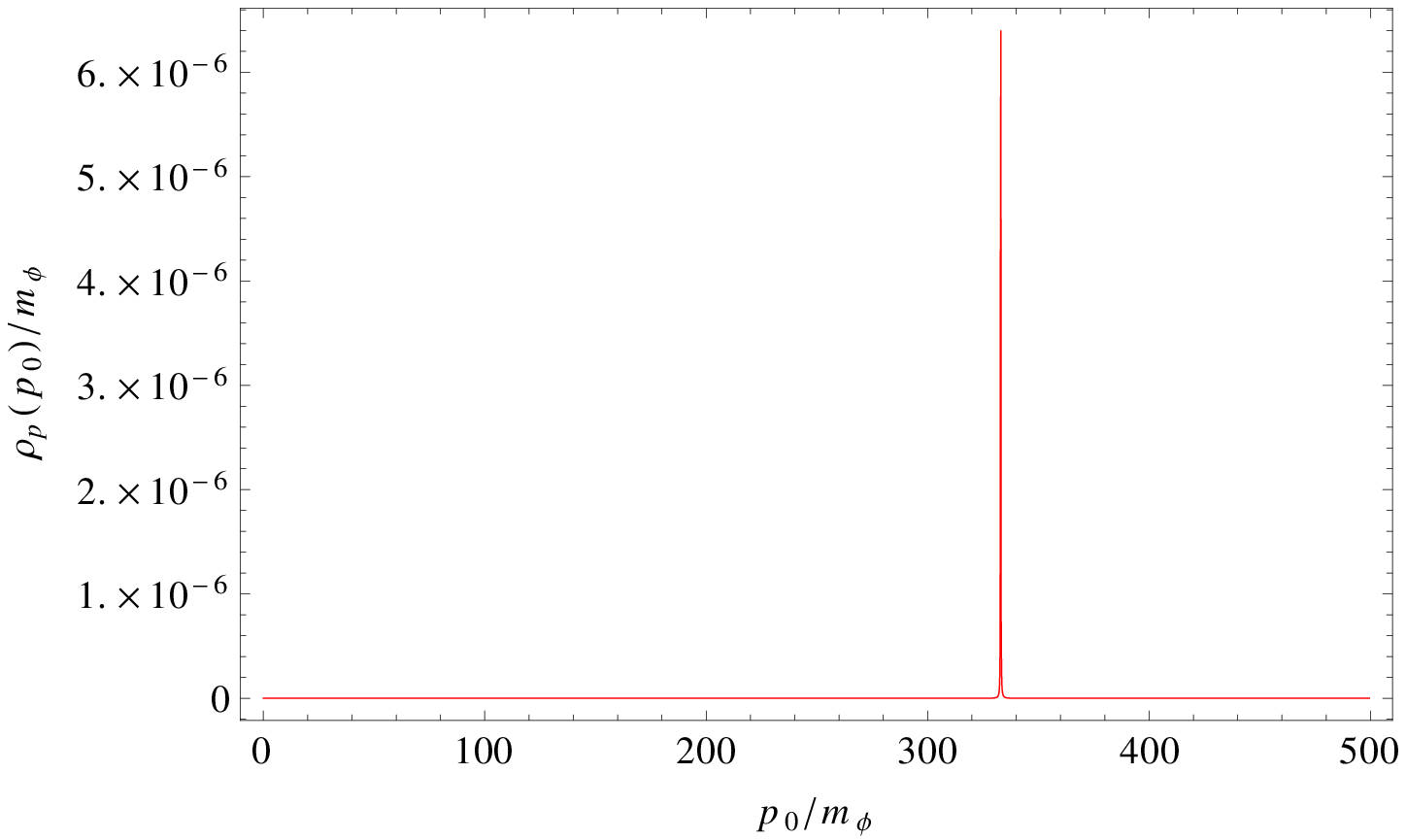}
    \caption{The spectral density $\rho_{\chi\p}(p_0)$ for $\chi$ with $g=m_\chi=m_\phi/6$, $T=333m_\phi$ and $|\p|=T$ as a function of $p_0$. 
\label{rhoplotchiT}}
\end{figure}

We first discuss the $\chi$-spectral function for momenta $\k\sim m_\phi$.
The real and imaginary parts of the $\chi$-self energy $\Uppi^R_{\chi\p}(p_0)$ for $|\k|=m_\phi$ are shown in figures \ref{RePiPlotchi} and \ref{ImPiPlotchi}. Figure \ref{rhoplotchi} shows the $\chi$-spectral density, obtained by inserting the above results for $\Uppi^{R}_{\chi,\k}(k_0)$ into (\ref{spectralfunction2}). 
The quasiparticle spectrum can be studied by solving the equation
\begin{equation}\label{dispersionrelationchi}
k_0^2-\k^2-m_\chi^2-{\rm Re}\Uppi^R_{\chi\k}(k_0)=0.
\end{equation}
For the parameters we considered it consists of a single peak, which is located slightly below the free particle energy $(\k^2+m_\chi^2)^{1/2}$ in vacuum, i.e. there is a small negative thermal mass shift. This mass shift, however, does not endanger the existence of the new scalar plasmon - if the diagram \ref{diagram}a) is evaluated with smaller $m_\chi$, then the spike in figure \ref{RePiPlot} moves to the left and remains in the regime where it can overcome $p_0^2-\p^2-m_\phi^2$ in (\ref{disprelfinder}). As for the screened $\phi$-particle, we observe a considerable thermal broadening. 
The sumrule (\ref{sumrule}) is fulfilled.
We can also calculate the $\chi$-spectral function for hard modes $\k\sim T$, which is shown in figure \ref{rhoplotchiT}.
As expected, it consists of one sharply defined  quasiparticle.
The screening leads to a small positive mass shift in this hard momentum region. 
This confirms that it is, up to finite width corrections, a reasonable approximation to use free thermal $\chi$-propagators when calculating the $\phi$-self-energy for loop momenta $\k\sim m_\phi$ and $\k\sim T$. This conclusion agrees with what we find in appendix \ref{DimensionalReduction}. 
Note that our analysis here goes beyond the hard thermal loop approximation, as (\ref{Uppi}) was obtained analytically without assumptions about the relative size of external and loop momenta.

The situation becomes more complicated when looking at $\chi$-modes with momenta $\k<m_\phi$, which also contribute to the $\chi$-self energy \ref{diagram}a).
Due to the smallness of $m_\chi$, it turns out that (\ref{dispersionrelationchi}) evaluated with (\ref{Uppi}) has no solutions at all for very small momenta.
One way to interpret this is that there are no $\chi$-quasiparticles with very small momenta, i.e. there is a minimum wave number below which the damping by the plasma is so strong that there are no plane waves that propagate for at least one oscillation. 
An interpretation of this is that $\chi$-particles travelling through the medium with very small momenta are ``halted'' by interactions with the background medium, which formally reflects in the ``melting'' of the quasiparticle peak for very soft modes at high $T$. 
If this is what physically happens, it could have considerable effect on the $\phi$-self-energy. The kinematic threshold in (\ref{xiIntSelfEn}), which is responsible for the appearance of the spikes in figure \ref{RePiPlot}, is caused by the energy conserving $\delta$-functions in (\ref{rhofree}).
As long as all widths remain narrow, thermal corrections encoded in (\ref{zerowidth}) will only move the mass shells and threshold. 
But if the peaks for soft modes indeed ``melt'', then (\ref{rhofree}) or (\ref{zerowidth}) by no means are a good approximation for the modes $\k,\p-\k \ll m_\phi$ in the loop. This could smear out the threshold in (\ref{xiIntSelfEn}) and hence the spike in figure \ref{RePiPlot}, which might eliminate the luon-solutions of (\ref{disprelfinder}) from the spectrum.
However, within our numerical precision the $\chi$-spectra for $|\k|\ll m_\phi$ obtained from (\ref{Uppi}) fail to satisfy the sum rule (\ref{sumrule}) with an error of order one.   
This indicates that our treatment of these modes is insufficient, and we can in fact not make any reliable statement on the $\chi$-quasiparticle spectrum at $\k\ll m_\phi$.

\subsection{Higher order corrections}\label{loopundso}
The problem is that we obtain the quasiparticle spectra from the diagrams in figure \ref{diagram}, the evaluation of which already requires knowledge of the full propagators (and thus the quasiparticle spectrum) in the loop.
In principle corrections to $\Pi^R_{\phi,\p}(p_0)$ are of higher order in the loop expansion and involve more vertices, hence one could expect them to be suppressed by powers of $g/m_\phi$.  
However, naive loop counting cannot always be applied in thermal field theory at high temperature because large occupation numbers can compensate for the suppression by additional vertices.
Formally this happens because the Bose-Einstein distribution in the thermal propagators can bring powers of the coupling into the denominator, so that the loop expansion need not be identical with the perturbative expansion. This problem is sometimes referred to as a ``breakdown of perturbation theory'' \cite{Linde:1980ts}. 

It can in some cases be fixed by using resummed propagators and vertices in the loops, see e.g. \cite{Kraemmer:2003gd}. 
Formally the resummed perturbation theory is nothing but a consistent perturbation theory, i.e. a systematic expansion in the coupling constant.\footnote{The need to use resummed propagators in loops can also be understood from simple kinematic arguments: If one uses the free thermal propagators given by (\ref{rhofree}) in the loop while assigning thermal masses to the external particle, then two propagators of the same particle with the same momentum meeting at a vertex would have different effective masses, which seems clearly unphysical.
%, which would imply that a particle can decay into itself and another particle in a two body decay. Such a decay may only be possible if the effective mass has strong momentum dependence, so that the decay products have a different effective mass.
} 
A scheme that is commonly used in gauge theories and for Yukawa interactions is the hard thermal loop (HTL) resummation technique \cite{Braaten:1989mz,Braaten:1990az,Frenkel:1989br}. 
The HTL resummation is a consistent expansion in a coupling constant $\alpha$ due to a separation of scales for $\alpha\ll1$. 
Thermal corrections to the dispersion relations of order $\sim\alpha T$ 
are relevant for soft external momenta $\p\lesssim \alpha T$. 
In contrast to that, the loop integral in the regime $T\gg m_\phi,m_\chi$ is usually dominated by hard loop momenta $\k\sim T$, for which thermal corrections are negligible. This justifies to use free thermal propagators inside the loop when calculating radiative corrections to propagators with soft external momenta.
The corrected propagators obtained this way can then be used in all further calculations. 

Unfortunately this strategy cannot be applied in the model (\ref{L}) because the self-energy diagram in figure \ref{diagram}a) for a particle with momentum $\p\ll T$ is, in contrast to gauge theories, not dominated by hard momenta $\sim T$ inside the loop. 
The reason is that the self-energies contain only scalar propagators. In comparison to, for instance, the fermion propagators appearing in the photon self-energy, these give less powers of momenta.\footnote{
Loop corrections have recently been studied beyond the HTL scheme in different contexts \cite{Drewes:2010pf,Anisimov:2010gy,Besak:2012qm,Garbrecht:2013gd,Garbrecht:2013bia,Laine:2013lka,Drewes:2013iaa,Salvio:2013iaa}, but these results from Yukawa and gauge theories cannot be applied to our model.
Resummations in scalar field theory have been studied in detail for the $\lambda\phi^4$-interaction, see e.g. \cite{Parwani:1991gq,Wang:1995qf,Aarts:1996qi,Buchmuller:1997yw,LeB,Kraemmer:2003gd,Drewes:2013iaa}. 
For that interaction considerable simplification arises from the fact that the leading order correction is a momentum-independent thermal mass of order $\sqrt{\lambda}T$ from a local diagram. This is not the case in the model (\ref{L}).}

Let $p$ be the external momentum of $\Pi^R_{\phi,\p}(p_0)$ and $k$ the momentum inside the loop.
If $\p$ is hard, i.e. comparable to $T$ or larger, then the thermal correction to the dispersion relation is small compared $\p$. The same argument justifies the use of free propagators in the part of the loop integration volume where $\k$ is hard. This allows to treat the effect of hard loops by means of resummed perturbation theory.
The uncertainty in our calculations of $\Pi^R_{\phi,\p}(p_0)$ arises from the region where $\p$ and $\k$ are both small compared to $T$ and $m_\phi$, i.e. from the interaction of the low momentum modes with each other.   
Similar uncertainties also exist in gauge theories, but there the loop integrals are usually dominated by the region $\k\sim T$, hence they only affect a sub-dominant contribution to the self-energy. 
In the scalar model under consideration here, in contrast, the integration region $\k\ll T$ considerably contributes to $\Pi^R_{\phi,\p}(p_0)$.
A fully consistent treatment of this issue goes beyond the scope of the current work.
Our simple estimate in appendix \ref{DimensionalReduction} suggests that thermal mass corrections to low momentum modes remain sufficiently small in the parameter region we consider. 
We postpone a more precise study to future work.

\section{Summary and Conclusion}
We have studied the spectrum of quasiparticles in a scalar model at high temperature at one loop level.
Our results indicate the existence of novel quasiparticles related to collective excitations in addition to the screened one-particle state for momenta $\p\lesssim m \ll T$, which we refer to as \emph{luons}.
The existence of purely collective excitations is in principle expected in any field theory at finite temperature at least for very long wavelength, where hydrodynamic modes contribute to transport. 
In gauge and Yukawa theories collective excitations are known to exist for soft momenta, i.e. much above regime where hydrodynamic excitations exist.
To the best of our knowledge, this phenomenon has not been described in a purely scalar field theory to date.

The new resonances lie in the region of a kinematic threshold, hence their properties may be sensitive to higher order corrections. 
We postpone the technically difficult treatment of these to future work.
If we assume that they do not considerably modify the luons or even eliminate them from the spectrum, then it is instructive to pose the question how general their appearance is. Mathematically they arise because the dispersive self-energy
exhibits a prominent negative spike that originates from the existence of a kinematic threshold 
in the dissipative self-energy 
at one-loop level.
It is related to the steep rise of (\ref{xiIntSelfEn}) in the threshold region. 
This suggests that luon-type plasmons can generally occur where 
the dissipative self-energy
is a steep function of $p_0$, such as near kinematic thresholds.
Hence, similar effects may also be expected in other, more realistic systems.

However, there are two generic properties of quasiparticle poles of this kind. 
First, since they occur due to a sharp feature or spike, they usually come with a suppressing residue factor $\mathcal{Z}_\p^{i}$  due to the steepness of 
the dispersive self-energy
in such a feature, see (\ref{ZandGamma}). That means that their contribution to phase space integrals is rather small, and the exchange of these quasiparticles will not contribute significantly to transport unless other channels are kinematically blocked. Such situations have been studied in cosmology \cite{Yokoyama:2005dv,Anisimov:2010gy,Drewes:2010pf,Kiessig:2010pr,Hamaguchi:2011jy,Drewes:2013iaa}.
Second, when their appearance is related to a kinematic threshold, they tend to lie in the threshold region, where changes in the shape of the dissipative self-energy
due to higher order corrections can have a strong effect on quasiparticle properties.

Finally, we would like to add that the collective excitations discussed here are absent in the most studied scalar model, the $\lambda\phi^4$-interaction. The reason is that 
the dispersive self-energy
in the $\lambda\phi^4$-model receives a momentum-independent contribution of order $\lambda T^2$ from a local diagram (``tadpole''). Any negative contribution to it, which could lead to additional poles in the spectral density, is of higher order in $\lambda$. This makes even the long wavelength modes too heavy for collective excitations of the type discussed here to occur.
\vspace{0.5cm}

{\large \textbf{Acknowledgements}} -
I am grateful to Mikhail Shaposhnikov, Dietrich B\"odeker and Miguel Escobedo for helpful comments. 
I would also like to thank Jin U Kang for pointing out that the result (\ref{Uppi}) can be written in a very compact form.
This work was supported by the Gottfried Wilhelm Leibniz program of the Deutsche Forschungsgemeinschaft.

\begin{appendix}
\section{Thermal mass and expansion parameter}\label{DimensionalReduction}
As discussed in section \ref{loopundso} it is not immediately clear that the loop expansion provides a systematic approximation scheme for small momenta in the model (\ref{L}).
We formulate an effective field theory for the low momentum modes, using the method of dimensional reduction \cite{Ginsparg:1980ef,Appelquist:1981vg,Farakos:1994kx,Farakos:1994xh,Kajantie:1995dw}. 
This procedure is commonly applied to static problems and in principle not suitable to treat dynamic quantities, such as the propagator we are interested here. Hence, we do not use it for a systematic treatment of the problem, but only to identify the relevant dimensionless expansion parameter.%\footnote{The situation in (massless) gauge theories is more clear because the coupling $\alpha$ is dimensionless,  i.e. $T$ is the only dimensionful parameter, hence for $\alpha\ll1$ all relevant scales can be distinguished as $\sim T$, $\sim\alpha T$, $\sim\alpha^2T$ etc.} 

Correlation functions for quantum fields in four dimensional Minkowski space at high temperature can be obtained by analytic continuation of correlators that have been calculated in a theory with three real spatial dimensions $\textbf{x}$ and an imaginary time dimension $i\tau$ on the compact interval from $0$ to $-i\beta$, with $\beta=1/T$ \cite{Matsubara:1955ws}. %\footnote{The reason is essentially that a thermal density matrix formally is a time evolution operator in imaginary time.}  
This leads to the Euclidean action
\begin{equation}\label{SE}
S_E=\int_0^\beta d\tau\int d^3\x \mathcal{L}_E
\end{equation}
with
\begin{equation}
\mathcal{L}_E= 
\frac{1}{2}(\partial_{\mu})^2\phi
+\frac{1}{2}m_{\phi,0}^{2}\phi^{2}
+\frac{1}{2}(\partial_{\mu}\chi)^2
+\frac{1}{2}m_{\chi,0}^{2}\chi^{2}
+g\phi\chi^2.
\end{equation}
Note that in this section $m_{\phi,0}$ and $m_{\chi,0}$ are to be understood as bare masses.
In this \textit{imaginary time formalism} the Fourier decomposition in energy is a discrete sum, 
\begin{eqnarray}
\phi(\x,\tau)=\sum_{n=-\infty}^{\infty}\phi_n(\x)e^{i\omega_n\tau}=\phi_0(\x)+\sum_{n=1}^{\infty}\left(\phi_n(\x)e^{i\omega_n\tau}+\phi_n^*(\x)e^{-i\omega_n\tau}\right),\label{phidecompo}\\
\chi(\x,\tau)=\sum_{n=-\infty}^{\infty}\chi_n(\x)e^{i\omega_n\tau}=\chi_0(\x)+\sum_{n=1}^{\infty}\left(\chi_n(\x)e^{i\omega_n\tau}+\chi_n^*(\x)e^{-i\omega_n\tau}\right),\label{chidecompo}
\end{eqnarray}
where we have used the reality of the fields in the second equality. Here $\omega_n\equiv2\pi Tn$, where $n$ is an integer.
Formally the four-dimensional theory with two fields $\phi(\x,\tau)$, $\chi(\x,\tau)$ is equivalent to an theory in three dimensional Euclidean space with infinitely many fields $\phi_n(\x)$, $\chi_n(\x)$. 
At high temperature modes with $n>0$ are heavy, and an effective field theory for modes $\p\ll 2\pi T$ can be constructed by integrating them out. The resulting effective Lagrangian is only valid for correlation functions with spatial separation $\ll 1/T$. Moreover, there is no more dependence on time, so one is restricted to static phenomena. %\footnote{The resulting two point function should be identified with the integral of the four dimensional correlator over $\tau$, i.e. for frequency zero.}.
We inserting (\ref{phidecompo}) and (\ref{chidecompo}) into (\ref{SE}) we can perform the $\tau$-integral 
analytically and write $S_E$ as a spatial integral over $\x$. 
Since we are only interested in the light modes $\phi_0$ and $\chi_0$ we drop all terms that do not contain these from $S_E$. In order to disentangle $\phi_n$ and $\phi_n^*$ we introduce new field variables
\begin{equation}
\rephi_n=\sqrt{\frac{2}{T}}{\rm Re}\phi_n \ , \ 
\imphi_n=\sqrt{\frac{2}{T}}{\rm Im}\phi_n \ , \
\rechi_n=\sqrt{\frac{2}{T}}{\rm Re}\chi_n \ , \
\imchi_n=\sqrt{\frac{2}{T}}{\rm Im}\chi_n 
.\end{equation}
We also define
\begin{equation}
\tphi\equiv\frac{\phi_0}{\sqrt{T}} \ , \ \tchi\equiv\frac{\chi_0}{\sqrt{T}} \ , \ \tg\equiv g\sqrt{T}.
\end{equation} 
With these new variables we can write
\begin{equation}
S_E=\int d^3\x \left[\tilde{\mathcal{L}} + \sum_{n=1}^{\infty} \mathcal{L}_n + \mathcal{L}_{\rm int}\right]
\end{equation}
with 
\begin{equation}
\tilde{\mathcal{L}}=\frac{1}{2}\Big((\partial_i\tphi)^2+m_{\phi,0}^2\tphi^2
+(\partial_i\tchi)^2+m_{\chi,0}^2\tchi^2\Big)
+\tg\tphi\tchi^2,
\end{equation}
\begin{eqnarray}
\mathcal{L}_n&=&\frac{1}{2}\Big((\partial_i\rephi_n)^2+(m_{\phi,0}^2+\omega_n^2 )\rephi_n^2
+(\partial_i\imphi_n)^2+(m_{\phi,0}^2+\omega_n^2)\imphi_n^2\nonumber\\
&&+(\partial_i\rechi_n)^2+(m_{\chi,0}^2+\omega_n^2 )\rechi_n^2
+(\partial_i\imchi_n)^2+(m_{\chi,0}^2+\omega_n^2)\imchi_n^2\Big),
\end{eqnarray}
\begin{equation}
\mathcal{L}_{\rm int}=\tg\sum_{n=1}^\infty\left[
2\tchi\left(\rephi_n\rechi_n+\imphi_n\imchi_n\right)
+\tphi\left(\rechi_n^2+\imchi_n^2\right)
\right]. 
\end{equation}
Correlation functions for $\tphi$ and $\tchi$ can be obtained from the generating functional
\begin{equation}\label{generating}
\mathcal{Z}=\int \mathcal{D}\tchi\mathcal{D}\tphi
(\prod_n\mathcal{D}\rechi_n\mathcal{D}\imchi_n\mathcal{D}\rephi_n\mathcal{D}\imphi_n
)
\exp
\left[
-\int d^3\x\left(\tilde{\mathcal{L}} + \sum_{n=1}^{\infty} \mathcal{L}_n + \mathcal{L}_{\rm int}
+J_{\tchi}\tchi + J_{\tphi} \tphi\right)
\right].
\end{equation}
We Taylor-expand the exponential in (\ref{generating}) to second order in $\mathcal{L}_{\rm int}$ and integrate out all fields except $\tphi$ and $\tchi$.  
This is equivalent to using the non-local Lagrangian
\begin{eqnarray}
\lefteqn{\frac{1}{2}\left((\partial_i\tphi(\x))^2+m_{\phi,0}^2\tphi(\x)^2
+(\partial_i\tchi(\x))^2+m_{\chi,0}(\x)^2\tchi^2\right)
+\tg\tphi(\x)\tchi(\x)^2}\nonumber\\
&&+\int d^3\y\left(
\tphi(\x)\mathcal{O}_{\tphi}(\x-\y)\tphi(\y)
+\tchi(\x)\mathcal{O}_{\tchi}(\x-\y)\tchi(\y)
\right)
\end{eqnarray}
with
\begin{eqnarray}
\mathcal{O}_{\tphi}(\x-\y)&=&\frac{1}{4}\tg^2\sum_{n=1}^{\infty} \rechi_n(\x)\rechi_n(\x)\rechi_n(\y)\rechi_n(\y)
,\end{eqnarray}
where we have used that the contributions from $\rechi_n$ and $\imchi_n$ are equal, and analogously
\begin{eqnarray}
\mathcal{O}_{\tchi}(\x-\y)&=&\tg^2\sum_{n=1}^{\infty} \rephi_n(\x)\rechi_n(\x)\rephi_n(\y)\rechi_n(\y)
.\end{eqnarray}
To obtain a local Lagrangian we can Fourier transform $\mathcal{O}_{\tphi}$ and $\mathcal{O}_{\tchi}$ and expand in momentum $\k$.
We keep only the $\k=0$ term. This is equivalent to using the effective Lagrangian 
\begin{eqnarray}
\mathcal{L}_{eff}=
\frac{1}{2}(\partial_i\tphi)^2
+\frac{1}{2}\tilde{m}_\phi^2\tphi^2
+\frac{1}{2}(\partial_i\tchi)^2
+\frac{1}{2}\tilde{m}_\chi^2\tchi^2
+\tg\tphi\tchi^2.
\end{eqnarray}
Here
\begin{eqnarray}
\tilde{m}_\phi^2&=&m_{\phi,0}^2+\frac{\tg^2}{2(2\pi)^d}\sum_n\frac{d^d\p}{\mu^{d-3}}\left( \p^2+m_{\chi,0}^2+\omega_n^2 \right)^{-2}\nonumber\\
&=&m_{\phi,0}^2+\frac{\tg^2}{2\mu^{d-3}(4\pi)^{d/2}}\frac{\Gamma(2-\frac{d}{2})}{\Gamma(2)}\sum_{n=1}^\infty(m_{\chi,0}^2+\omega_n^2)^{d/2-2}
\end{eqnarray}
We have used dimensional regularisation; though each term in the sum over $n$ is finite, the sum diverges. 
The origin is the usual UV-divergence in the four dimensional theory.
If we neglect the masses $m_{\phi,0}$, $m_{\chi,0}$ in the loop we can perform the sum,
\begin{eqnarray}
\tilde{m}_\phi^2
=m_{\phi,0}^2+\frac{\tg^2}{2\mu^{d-3}(4\pi)^{d/2}}\frac{\Gamma(2-\frac{d}{2})}{\Gamma(2)}(2\pi T)^{d-4}\zeta(4-d).
\end{eqnarray}
Setting $d=3-\epsilon$ and expanding in $\epsilon$ we obtain
\begin{eqnarray}
\tilde{m}_\phi^2
=m_{\phi,0}^2
+\frac{g^2}{2(4\pi)^2}\left(
\frac{1}{\epsilon}
+\gamma_E
+\frac{1}{2}[\psi(1/2)-{\rm log}\pi] 
+{\rm log}\left[\frac{\mu}{T}\right]
\right)
\end{eqnarray}
where $\gamma_E\simeq0.5772$ and $\psi$ is the digamma function. 
Similarly we obtain 
\begin{eqnarray}\label{chimasscorrdimred}
\tilde{m}_\chi^2=m_{\chi,0}^2+
\frac{g^2}{(4\pi)^2}\left(
\frac{1}{\epsilon}
+\gamma_E
+\frac{1}{2}[\psi(1/2)-{\rm log}\pi] 
+{\rm log}\left[\frac{\mu}{T}\right]
\right)
\end{eqnarray}
The factor $2$ difference in the loop correction can be identified with the symmetry factor of the $\phi$-self-energy diagram.
We can easily relate the temperature independent terms in (\ref{chimasscorrdimred}) to the one-loop self-energy correction for vanishing external momentum in the four dimensional theory (\ref{L}) in vacuum, 
\begin{equation}\label{zeroT}
\frac{g^2}{(4\pi)^2}\left(
\frac{1}{\epsilon}
+\frac{\gamma_E}{2}
-{\rm log}(4\pi) 
+{\rm log}\left[\frac{m_\chi}{\mu}\right]
\right)
\end{equation}
We impose renormalisation conditions at $T=0$ and eliminate the logarithm in (\ref{zeroT}) by fixing fix $\mu=m$, leading to a finite temperature contribution
\begin{eqnarray}
\frac{g^2}{(4\pi)^2}{\rm log}\left[\frac{m_\chi}{T}\right]
,\end{eqnarray}
to $\tilde{m}_\chi$. There is an analogue correction
\begin{eqnarray}
\frac{g^2}{2(4\pi)^2}{\rm log}\left[\frac{m_\phi}{T}\right]
\end{eqnarray}
to $\tilde{m}_\phi$.
This shows that indeed the mass correction from heavy modes (hard thermal loops) is small, 
and we can confirm that hard modes can be treated perturbatively for our choice of parameters. 
To estimate the effect of the light modes onto themselves, let us now consider a loop correction within the effective theory. At vanishing external momentum we formally obtain a correction 
\begin{equation}\label{largemass}
\sim\frac{g^2T}{4\pi(\tilde{m}_\phi+\tilde{m}_\chi)}
\end{equation}
to $\tilde{m}_\chi$, which can exceed $m_\chi$ by a factor of a few due to the growth of the effective coupling $\tilde{g}=g\sqrt{T}$ with temperature, but remains small compared to $m_\phi$ for our choice of parameters. 
This suggests thermal masses remain sufficiently small that there exist choices of vacuum masses for which the novel collective excitations we found exist. 
A fully consistent treatment would, however, require more refined calculations. 
\end{appendix}

\bibliographystyle{JHEP}
\bibliography{all}
\end{document}